\documentclass[preprint,12pt]{elsarticle}
\usepackage{graphicx}
\usepackage{amsmath,amsthm,amssymb}
 
\usepackage{bm}
\usepackage{color}

\newtheorem{proposition}{Proposition}

\newcommand{\beq}{\begin{equation}}
\newcommand{\beqa}{\begin{eqnarray*}}
\newcommand{\beqan}{\begin{eqnarray}}
\newcommand{\greq}{\begin{equation}\left\{ \begin{array}{l}}
\newcommand{\eeq}{\end{equation}} 
\newcommand{\eeqa}{\end{eqnarray*}}
\newcommand{\eeqan}{\end{eqnarray}}

\newcommand{\lp}{ \left(}
\newcommand{\rp}{ \right)}

\begin{document}

\newcommand{\ddd}[2]{\frac{\delta  #1}{\delta #2}}  
\newcommand{\ddx}[1]{\frac{\partial }{\partial #1}}  

\def\MM{{\cal M}}
\def\NN{{\cal N}}

\def\R{I\!\!R}
\def\L{I\!\!L^2}

 \begin{frontmatter}
 
\title{The Hamiltonian description  of incompressible fluid ellipsoids}

\author{P.J.~Morrison\corref{cor1}}
\ead{morrison@physics.utexas.edu}
\address{Department of Physics and Institute for Fusion Studies, 
The University of Texas at Austin, Austin, TX, 78712, USA}
\author{Norman R.~Lebovitz}
\address{Department of Mathematics, The University of Chicago,
        Chicago, IL, 60637, USA}
\author{Joseph A.~Biello}
\address{Department of Mathematics, University of California at Davis, USA}

\cortext[cor1]{Correponding author}

\begin{abstract}

We construct the noncanonical Poisson bracket associated with the phase space of
first order moments of the velocity field and quadratic moments of the
density of a fluid with a free-boundary, constrained by the condition
of incompressibility. Two methods are used to obtain the bracket, both based on Dirac's
procedure for incorporating constraints. First, the Poisson bracket of
moments of the unconstrained Euler equations is used to construct a
Dirac bracket, with Casimir invariants corresponding to volume
preservation and incompressibility.  Second, the Dirac procedure is
applied directly to the continuum, noncanonical Poisson bracket that describes
the compressible Euler equations, and the moment reduction is applied
to this bracket.  When the Hamiltonian can be expressed exactly in
terms of these moments, a closure is achieved and the resulting
finite-dimensional Hamiltonian system provides exact solutions of
Euler's equations.  This is shown to be the case for the classical,
incompressible Riemann ellipsoids, which have velocities that vary
linearly with position and have constant density within an ellipsoidal
boundary.  The incompressible, noncanonical Poisson bracket differs
from its counterpart for the compressible case in that it is not of
Lie-Poisson form.

\end{abstract}


 \end{frontmatter}

\baselineskip 24 pt 


\section{Introduction}

The Euler equations governing the velocity field ${\bf v}$, density $\rho$ and pressure $p$ of an inviscid fluid are
\begin{eqnarray}\label{euler1}
\partial_t{\bf v} + {\bf v}\cdot\nabla{\bf v} &=&
 -\rho ^{-1}\nabla p + {\bf f} \;\quad \mbox{ and}   \\
\partial_t{\rho} + \mbox{div}\, \left( \rho  {\bf v} \right)  &=& 0
\label{euler2}
\end{eqnarray}
where ${\bf f}$ denotes an as yet unspecified force. These equations
must be augmented by boundary and initial data, and by further
conditions relating the variables ${\bf v}, p \; \mbox{and} \; \rho$: either an
equation of state (in the compressible case) or the condition
$\mbox{div}\, {\bf v}=0$ (in the incompressible case). We shall be
interested in exploring their Hamiltonian structure in a particular
context. Our principal reference for a general discussion of this
structure and the derivations of the corresponding brackets will be
\cite{morrison98}.

Exact solutions of the Euler equations are possible only under
simplifying assumptions and in simple contexts. A family of solutions
in the context of astrophysics, namely, where the force term ${\bf f}$
includes the self-gravitational effects of the fluid mass, exists
under the assumption of a fluid of uniform density confined to an
ellipsoidal domain, with a velocity field linear in the
coordinates. These assumptions reduce the Euler equations to a finite
system of ordinary differential equations. The equations for these
{\em Riemann ellipsoids} have been widely investigated: their study
goes back to the work of Dirichlet (\cite{dirichlet}) and of Riemann
(\cite{riemann}), but our principal reference for this will be
\cite{chandra}. We summarize their properties in Appendix
\ref{summary}, which we will refer to often.  (In Appendix \ref{hybrid} we describe four natural frames of reference for the ellipsoids, which are included here because they do not appear to have been published together elsewhere.)

In none of these references was the Hamiltonian nature of the
finite-dimensional system emphasized. This was first addressed  by
Rosensteel \cite{rosensteel88}. His starting point was the so-called {\em virial}
method originally introduced to investigate the stability of steady
solutions of the Euler equations. The {\em virial} is a moment of the
form
\begin{equation}
\label{defmom1}
{\cal M}_{ij}= \int _D \rho x_i v_j \, d^3x=\int _D x_i M_j \, d^3x,
 \end{equation} 
where $i,j =1,2,3,$ and the second form introduces the
specific momentum of the fluid ${\bf M}=\rho {\bf v}$.  This moment is considered together
with another moment, equivalent to the moment-of-inertia tensor,
\begin{equation}
\label{defmom2}
\Sigma _{ij}= \int _D \rho x_i x_j \, d^3x \;. 
\end{equation} 
Rosensteel presents an algebra for these
moments, i.e., bracket relations among them that are closed, and that
provide a noncanonical Hamiltonian description of the Riemann
ellipsoids with a certain choice of the Hamiltonian function
$H(\Sigma, {\cal M})$; we present these relations below, in equations
(\ref{S_S_lp})-(\ref{M_S_lp}).

We call attention to two features of Rosensteel's description of the
incompressible case:
\begin{enumerate}
\item \label{fd_rel} The bracket relations are presented without
reference to the fluid-dynamics equations (\ref{euler1}) and
(\ref{euler2}) above, and
\item \label{inc_cond} The formulation requires a Hamiltonian function
other than the total energy as well as the imposition of extraneous
constraints.

\end{enumerate}

The feature (\ref{fd_rel}) is addressed in \S \ref{lp} below, where we
derive Rosensteel's bracket relations in a straigtforward way via a
moment reduction of the general fluid-dynamical bracket
(\ref{lie-poisson2}). Feature (\ref{inc_cond}) is discussed in detail
in \S\S \ref{rosensec} and \ref{diracsec}.

We view the fluid as incompressible. This is natural because the
density of the Riemann ellipsoids is spatially uniform\footnote{There
are also applications allowing for compressibility wherein $\rho =
\rho (t)$, i.e., the density is spatially uniform but varies with
time. We do not address these cases here.}. However, Rosensteel's
bracket does not constrain the fluid to be incompressible, and we
therefore modify it via Dirac's procedure for incorporating
constraints. Dirac's method is described in \S \ref{dbmom}. We observe
in \S \ref{dbflu} that one can alternatively first apply Dirac's
procedure and subsequently effect a moment reduction, with the same
result. The resulting Dirac
bracket is no longer of Lie-Poisson type: the bracket relations depend
nonlinearly on the moments.

In \S \ref{rosensec} we relate the noncanonical Hamiltonian equations
obtained from Rosensteel's bracket to the equations describing the
Riemann ellipsoids and show that, if the Hamiltonian is taken to be
the total energy, the pressure term from fluid dynamics is missing. It
can be restored by adding an extra term to the Hamiltonian. In \S
\ref{diracsec} we show that the Hamiltonian equations obtained from
the Dirac bracket using the total energy as Hamiltonian give the full
equations for the Riemann ellipsoids, and, moreover, avoid the
necessity of imposing any further constraints. Finally, in \S \ref{disc} 
we summarize and discuss these results. 


\section{The Lie-Poisson Bracket and its Moment Reduction}
\label{lp}

The Euler equations (\ref{euler1}) and (\ref{euler2}) can be
re-expressed in terms of the momentum density, ${\bf M} := \rho {\bf
v}$, as
\beqan
\partial_t{\bf M} +  {\bf v}\cdot \nabla {\bf M} + \left(
\mbox{div}\,{\bf v}\right) {\bf M}&=& - \nabla p + \rho
{\bf f} \; \mbox{and}\label{M_eq}\\
\partial_t\rho +  \mbox{div}\,{\bf M} &=& 0 \label{rho_eq}.
\eeqan
These, like equations (\ref{euler1}) and (\ref{euler2}), will be referred to as
unconstrained, since neither the constraint of incompressibility nor
that of an equation of state has yet been imposed.

The    Hamiltonian description of these equations is reviewed in \cite{morrison98}.  The noncanonical Poisson bracket, as given in \cite{morrison80}, is\footnote{Unless
otherwise indicated, repeated Latin indices are summed from 1 to 3.}
\beqan
\{F,G\}_M &=&  \int_{{\R}^3} M_i 
\left( \ddd{G}{M_j} \frac{\partial}{\partial x_j} \ddd{F}{M_i} - 
\ddd{F}{M_j} \frac{\partial}{\partial x_j} \ddd{G}{M_i}
\right)  \,\,\,d^3 x  \nonumber \\
&+& \int_{{\R}^3} \rho \left(
\ddd{G}{{\bf M}} \cdot {\bf \nabla}\, \ddd{F}{\rho}  -
\ddd{F}{{\bf M}} \cdot {\bf \nabla}\, \ddd{G}{\rho}  
\right) \,\,\,d^3 x. 
\label{lie-poisson2}
\eeqan
This is a Lie-Poisson bracket (i.e., is linear in the variables ${\bf
M}$ and $\rho$). It is implicit in the derivation of this bracket that
the integrals are convergent, i.e., the density and momentum
variables, and the functions of them that appear in the integrals,
fall off sufficiently fast at large distances. The subscript $M$
indicates that this version employs the momentum (as opposed to the
velocity) as a dynamical variable. Since several brackets appear
below, we'll use subscripts to distinguish among them. This bracket,
like the versions of the Euler equations given above, is
unconstrained. It allows for compressiblity, which is, however,
expressed explicitly only in the Hamiltonian:
\beq
H = \int_{{\R}^3} \left(\frac{|{\bf M}|^2}{2 \rho} + \rho U(\rho) +
{\rho} \chi \right) d^3x ,
\eeq
where ${\bf f}= \nabla \chi$ and $U$ represents the internal energy.
This bracket and this Hamiltonian generate the compressible Euler
equations with the pressure given by
\beq
p = \rho^2 \frac{\partial U}{\partial \rho}\,.
\eeq

We can apply the  bracket (\ref{lie-poisson2}) above to the functionals ${\cal M}_{ij},
\Sigma _{ij}, i,j=1,3$. Since (see
equations \ref{defmom1} and \ref{defmom2} above)
\begin{eqnarray}
&& \frac{\delta {\cal M}_{ij}}{\delta M_k}=x_i \delta _{jk} \quad  \mbox{and}\quad  \frac{\delta
{\cal M}_{ij}}{\delta \rho}=0 ;\label{mom1_fds}\\
&&  \frac{\delta \Sigma_{ij}}{\delta M_k}=0 \quad \mbox{and}\quad  \frac{\delta
\Sigma_{ij}}{\delta \rho}=x_ix_j    ,\label{mom2_fds}
\end{eqnarray}
we easily find the following bracket relations:

\beqan
\{\Sigma_{ij},\Sigma_{kl}\}_{R} &=& 0
\label{S_S_lp}\\
\{\MM_{ij},\MM_{kl}\}_{R}
&=& \delta_{il} \MM_{kj} - \delta_{jk} \MM_{il} 
\label{M_M_lp} \\
\{\Sigma_{ij},\MM_{kl}\}_{R} &=&
  \delta_{il} \Sigma_{jk} + \delta_{jl} \Sigma_{ik}.
\label{M_S_lp} 
\eeqan
These are precisely the relations obtained by Rosensteel by other means
(\cite{rosensteel88}, eq. 134); hence the index $R$. Since they are
obtained directly from the unconstrained bracket (\ref{lie-poisson2}),
they must be likewise unconstrained.


\section{Dirac bracket for the Moment Formulation}
\label{dbmom}

We address here the constraint of incompressibility, which is not
incorporated in the bracket relations (\ref{S_S_lp})-(\ref{M_S_lp})
above. We do this with the aid of the Dirac-bracket formalism. We
begin this section by defining this and conclude by giving the
relations (\ref{S_S_D}), (\ref{M_M_D}) and (\ref{M_S_D}) for the Dirac
bracket obtained from Rosensteel's bracket.

\subsection{The Dirac Bracket}
\label{dbrac}

Given a bracket $\{\cdot \, , \, \cdot \}$, canonical or noncanonical,  and an even number $2k$ of phase-space functions $\{C^{\mu}\}_1^{2k}$, one can define a new
bracket for which these functions are Casimir invariants. 
This (so-called) Dirac bracket is
constructed as follows:
\beq
\{F,G\}_D = \{F,G\} - \sum _{\mu ,\nu=1}^{2k} \{F,C^{\mu}\} \,\omega^{-1}_{\mu \nu} \,
\{C^{\nu},G\} 
\label{dirac}
\eeq
where
\beq
\omega^{\mu \nu} = \{C^{\mu},C^{\nu}\}
\label{dirac_omega};
\eeq
it is further assumed that $\omega$, an  antisymmetric matrix function of the dynamical
variables, is  invertible. The following observations follow directly from this
definition:
\begin{enumerate}
\item Each of the functions $C^\mu$ is a
Casimir invariant for the Dirac bracket
\item Any Casimir invariant of the original bracket $\{\cdot \, , \,
\cdot \}$ is likewise a Casimir invariant for $\{\cdot \, , \,
\cdot\}_D$.
\item  $\{\cdot \, , \, \cdot \}_{D}$ is
antisymmetric and satisfies the Leibnitz rule (as in equation
\ref{leibnitz} below).
\end{enumerate}
It is less obvious but also true that it satisfies the
Jacobi identity. This is proved in Appendix \ref{jacobi} below.

Suppose now that $P$ is a constant of the motion in the dynamics
provided by a particular Hamiltonian function $H$ under the original
bracket, but not a Casimir: $\{P,H\}=0$ but $\{P,G\} \ne 0$ for some
phase-space function $G$. Then it is not guaranteed that $P$ will be
a constant of the motion in the dynamics provided by $H$ under the
modified bracket $\{\cdot \, , \, \cdot\}_D$; it's possible in
principle that $\{P,H\}_D \ne 0$. An example of this is given in
Appendix \ref{example}. However, {\em some} constants of the motion $P$ remain
constants of the motion under the modified bracket. The following
proposition is easily verified:
\begin{proposition}\label{little}
If $P=H$, the Hamiltonian, or if $\{P,C^\mu \}=0$ for each $\mu = 1,2, \ldots ,2k,$ then
$P$ is a constant of the motion in the dynamics provided by $H$ also
under the Dirac bracket $\{ \cdot \, , \, \cdot \}_D$.
\end{proposition}

In the application of the present paper we find that the constants of
the motion are in fact unchanged. We have $k=1$ and the functions $C^1$ and
$C^2$ are given by equation (\ref{C1C2defs}) below. The only constants of
the motion that are not Casimirs of the original bracket are the
Hamiltionian $H$ and the three components of the angular momentum
\[ L_i = \epsilon _{ijk} {\cal M}_{jk}, i=1,2,3.\]
These commute with $C^1$ and $C^2$ by virtue of the formulas
(\ref{sigmaC}) and (\ref{MC}) below. The persistence of the constants of the
motion under the change of bracket follows therefore from the
Proposition (\ref{little}).

\subsection{A Pair of Constraints}
\label{constraints}

We choose for the original bracket that of Rosensteel, whose
relations are given in equations (\ref{S_S_lp})-(\ref{M_S_lp}) above. 

As discussed in \S \ref{disc} below, this bracket has a Casimir whose
fluid-dynamical interpretation is the magnitude of the circulation
vector.  In order to construct the incompressible bracket we shall
augment this algebra by adding two additional Casimirs $C^1$ and $C^2$
expressing the constancy of the volume and constancy of the divergence
of the velocity field. We may express these in the forms
\beq \label{C1C2defs}
C^1 = \ln \lp{\rm Det}\left(\Sigma\right)\rp \; \mbox{and} \; C^2 =
{\rm Tr}\left(\Sigma^{-1}{\cal M}\right). \eeq 
The explanation for these choices originates
in the context of a fluid confined to an ellipsoidal domain and having
velocity components that are linear in the cartesian
coordinates. Consider $C^1$ first. The moment tensor $\Sigma$ is
symmetric and, when transformed to a principal-axis frame for the
ellipsoid, takes the form
\beq 
\label
{Q} Q =(m/5)\mbox{diag}\left(a_1^2,a_2^2,a_3^2 \right)\eeq
where $a_1,a_2$, and $a_3$ are the principal-axis lengths and  $m$ is the total mass\footnote{The total mass is conserved; it is a
Casimir invariant of the bracket (\ref{lie-poisson2}) (see \cite{morrison98}).}. Therefore
\beq 
\mbox{det}(\Sigma)=(m/5)^3 \left(a_1a_2a_3\right)^2
\label{detsigma}
 \eeq
and $C^1$ as defined above is a constant of the motion as long as the volume $(4/3)\pi
a_1a_2a_3$ is. Since for a figure of uniform density the
constancy of the volume implies that of the density, the constancy of
$C^1$ can be viewed equally as the constancy of the density $\rho$.
Regarding $C^2$,  we note that for a fluid having a linear velocity
field $V=L(t)X$ for some matrix $L$, the divergence of the velocity is
the trace of $L$, which should therefore vanish under the assumption
of incompressibility. Substituting the expression for $V$ into the
moment equation (\ref{defmom1}), we find that $L^t=\Sigma ^{-1}{\cal
M}$. Therefore
\[ \mbox{Tr}(L)=\mbox{Tr}(L^t)=Tr(\Sigma ^{-1}{\cal M})\]
and the velocity field is solenoidal if $C^2 =0.$

\subsection{Some Useful Formulas}
\label{formulas}

The calculation of the Dirac-bracket relations and of other related
quantities needed below requires some preliminary formulas, which we
record here. Two useful, general identities for matrices $A=(A_{ij})$
are
\[
 \frac{\partial A^{-1}_{ij}}{\partial A_{kl}}=-A^{-1}_{ik}A^{-1}_{lj}
\quad \mbox{and} \quad \frac{\partial \mbox{det}A}{\partial A_{ij}}=
C_{ij}\,,
\] 
where $C_{ij}$ is the cofactor of $A_{ij}$.

In order to apply the bracket relations to arbitrary functions of
functionals, we use the derivative propery of brackets: if
$v_1,v_2,\ldots v_k$ are functionals and $g(v)=g(v_1,v_2, \ldots
,v_k)$ is a real-valued function of them, then for any other
functional $u$
\beq \label{leibnitz} \{u,g(v)\}= \sum _{i=1}^k \frac{\partial g}{\partial v_i} \{u,v_i\}.\eeq
We can now record the
following relations for Rosensteel's bracket:
\beqan \label{Crels}
\left\{ \Sigma_{ij}, C^1 \right\}_R &=& 0 \quad \mbox{and} \quad
\left\{ \Sigma_{ij}, C^2 \right\}_R = 2 \delta_{ij} ,\label{sigmaC}\\
\left\{ \MM_{ij}, C^1 \right\}_R &=& -2 \delta_{ij} \quad \mbox{and}
\quad \left\{ \MM_{ij}, C^2 \right\}_R = \Sigma^{-1}_{in} \MM_{nj} +
\Sigma^{-1}_{jn} \MM_{ni} \label{MC}
\eeqan

\subsection{Dirac Bracket for Incompressible Ellipsoids}
\label{rebrac}

Since there are only two constraints, the matrix $\omega$ has only one  independent entry, 
\beqan
\omega^{1 \,2} &=& -\omega^{2 \,1} = \left\{C^1,C^2\right\}_R = \Sigma^{-1}_{kl} \left\{  \Sigma_{kl} , \MM_{ij} \right\}
\Sigma^{-1}_{ij} \nonumber \\ 
&=& \Sigma^{-1}_{ij} \lp \delta_{jk} \Sigma_{il} + \delta_{jl} \Sigma_{ik} \rp
\Sigma^{-1}_{kl} = 2 {\rm Tr}\left(\Sigma^{-1} \right) \label{om_proj}
\eeqan
which implies
\beq
\omega^{-1} = \frac{-1}{2 {\rm Tr}\left(\Sigma^{-1} \right)}\left[
  \begin{array}{cc} 0 & 1 \\ -1 & 0 
  \end{array} \right] .
\eeq

Thus, the relations for the Rosensteel-Dirac bracket become
\beqan
\{\Sigma_{ij},\Sigma_{kl}\}_{RD} &=& 0,
\label{S_S_D}\\
\{\MM_{ij},\MM_{kl}\}_{RD} &=& \{\MM_{ij},\MM_{kl}\}_R 
\nonumber \\
&+&   \frac{1}{\mbox{Tr}(\Sigma ^{-1})} \left(  \delta_{ij} \lp \Sigma^{-1}_{kn} \MM_{nl} +
         \Sigma^{-1}_{ln}  \MM_{nk}  \rp
    -  \delta_{kl} \lp \Sigma^{-1}_{in} \MM_{nj} +
    \Sigma^{-1}_{jn}  \MM_{ni}    \rp \right)\;\mbox{and}
\label{M_M_D} \\ 
\{\MM_{ij},\Sigma_{kl}\}_{RD} &=& \{\MM_{ij},\Sigma_{kl}\}_R + \frac{2 \delta_{ij} \delta_{kl}}{{\rm
    Tr}\left(\Sigma^{-1} \right). }  
\label{M_S_D}
\eeqan

These bracket relations are nonlinear, i.e., the Dirac bracket is no
longer of Lie-Poisson type. We return later (\S \ref{diracsec}) to a verification that
they provide a Hamiltonian description of the equations governing the
motions of the incompressible Riemann ellipsoids.


\section{Dirac Bracket for the Fluid Formulation}
\label{dbflu}

We obtain a different route to the Dirac bracket for the
incompressible, Riemann ellipsoids if we first constrain
the fluid bracket (\ref{lie-poisson2}) and only subsequently perform
the moment reduction. For this purpose we carry out the procedure
embodied in equation (\ref{dirac}) but for the original bracket we
employ the fluid-dynamical bracket (\ref{lie-poisson2}). We impose
the same constraints $C^1$ and $C^2$ as defined in equation
(\ref{C1C2defs}) and therefore need expressions for the brackets
$\{F,C^1\}_M,\{F,C^2\}_M$ and $\{C^1,C^2\}_M$. For this we need
the variational derivatives of $C^1$ and $C^2$ with
respect to the variables ${\bf M}$ and $\rho$. Straightforward
calculations lead to the following:
\beqan
\frac{\delta C^1}{\delta M_k}= 0 \quad & \mbox{and} & \quad \frac{\delta
C^1}{\delta \rho}= \Sigma ^{-1}_{ij}x_ix_j \label{deltaC1}; \\
\frac{\delta C^2}{\delta M_k}=\Sigma
^{-1}_{kj}x_j  \quad & \mbox{and} & \quad
\frac{\delta C^2}{\delta \rho}= -{\cal M}_{kl}\Sigma ^{-1}_{ik}\Sigma
^{-1}_{jl}x_ix_j.  \label{deltaC2}
\eeqan

The expressions needed for modifying the bracket are easily obtained with
the aid of equations (\ref{deltaC1}) and (\ref{deltaC2}):
\beqan
 &&\left\{F,C^1\right\}_M=-2 \Sigma ^{-1}_{lm} \int _{R^3} \rho x_m
\frac{\delta F}{\delta M_l}\,d^3x, 
\label{FC1} \\
&& \left\{F,C^2\right\}_M=\int _{R^3} M_i\left[\Sigma ^{-1}_{jl}x_l
\frac{\partial}{\partial x_j} \frac{\delta F}{\delta M_i} - \Sigma
^{-1}_{ij} \frac{\delta F}{\delta M_j} \right]\, d^3x \nonumber \\
&+&  \int _{R^3} \rho \left[ \Sigma ^{-1}_{kj}x_j \frac{\partial }{\partial x_k}
\frac{\delta F}{\delta \rho} + x_i \frac{\delta F}{\delta
M_j}\left(A_{ji} + A_{ij}\right)  \right] \, d^3x,\label{FC2}
\eeqan
where $A= \Sigma^{-1}{\cal M} \Sigma ^{-1}$;
and, from either of the preceding equations,
\beq 
\label{C1C2}
\left\{C^1,C^2\right\}_M=2 \mbox{Tr}\,\left(\Sigma ^{-1} \right).
\eeq

\begin{figure}
%
\centering \includegraphics[width=3in]{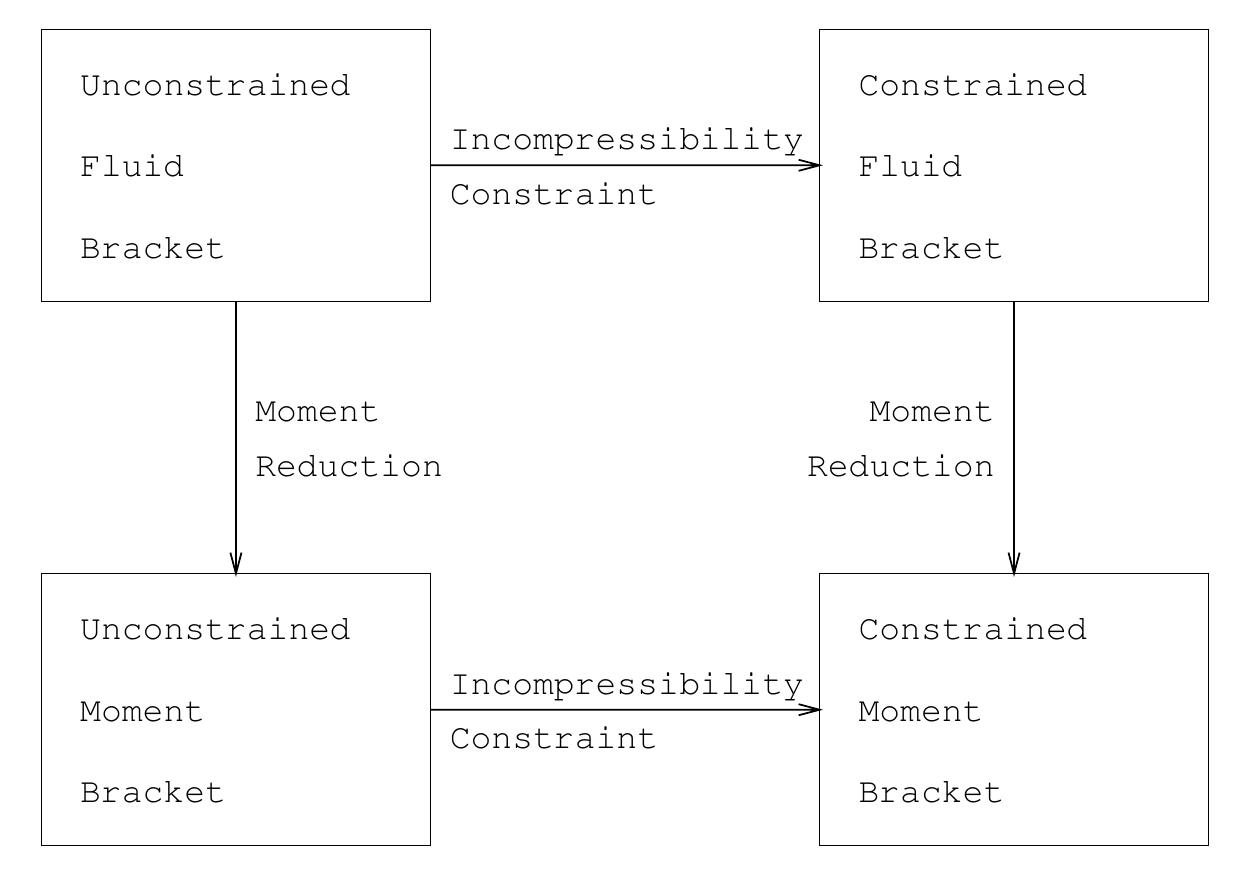}
\caption{\small This commuting diagram indicates that the final result
is obtained on following either path.}
\label{cd}
\end{figure}

The Dirac-constrained fluid bracket is therefore
\beq
 \left\{F,G\right\}_{MD}=\left\{F,G\right\}_M +
 \frac{1}{2\mbox{Tr}\left(\Sigma ^{-1}\right)} \left(
\left\{F,C^1\right\}_M \left\{ C^2,G\right\}_M - \left\{F,C^2
\right\}_M \left\{ C^1,G \right\}_M\right),\label{MDbracket}\eeq
where the index $MD$ denotes the momentum-Dirac bracket, and the index $M$
denotes the unconstrained fluid bracket
(\ref{lie-poisson2}). We next carry out the moment reduction with the
aid of equations (\ref{mom1_fds}) and (\ref{mom2_fds}). We find
\beqan 
\left\{\Sigma _{ij},C^1\right\}_M=0  &\quad \mbox{and} \quad& \left\{\Sigma
_{ij},C^2\right\}_M = 2\delta _{ij}; \label{sigmoms}\\
 \left\{ {\cal M}_{ij},C^1\right\}_M = -2\delta _{ij} &\quad \mbox{and} \quad&
\left\{ {\cal M} _{ij},C^2\right\}_M=\Sigma ^{-1}_{ik} {\cal M}_{kj} + \Sigma ^{-1}_{jk} {\cal M}_{ki}. 
\label{mmoms}
\eeqan
These are exactly the same expressions as found in the preceding
section where the braces referred to the finite system of
Rosensteel's relations (\ref{S_S_lp}), (\ref{M_M_lp}) and
(\ref{M_S_lp}). Since the moment reduction of the first term on the
right-hand side of equation (\ref{MDbracket}) leads as we have seen to
Rosensteel's bracket, we arrive at the same constrained, moment bracket
via either route, as indicated in Figure \ref{cd}.

In the next two sections we investigate the structure of Hamilton's equations first using
Rosensteel's bracket and then using the Dirac bracket based on it.   


\section{Dynamical Equations Under Rosensteel's Bracket}
\label{rosensec}

In this section we work out the dynamical equations obtained under Rosensteel's bracket, using as the Hamiltonian function the total energy of a Riemann
ellipsoid. We shall find (see the last sentence of this section) that a key term is missing. 

The symmetric matrix $\Sigma$ can be
transformed to the diagonal form $Q$, as in equation (\ref{Q}) above. We
have implicitly assumed in this description that $\Sigma$ is positive-definite: this represents a
choice of initial data and, once made, will persist for at least a
finite time interval. We
assign the potential energy appropriate to an ellipsoid with semiaxes $a_1,a_2,a_3$:
\beq
 \label{poten}
{\cal W}= -(1/2)\int \rho V(x) d^3x = -(3/10)m^2G{\cal I},
\eeq
where the potential function $V$ is given by equation (\ref{gravpot})
and ${\cal I}$ by equation (\ref{coeffs}) of Appendix \ref{summary}.
We have further used equation (22) of chapter 3 of \cite{chandra} to
complete the integration. This potential energy is therefore a
function only of the squares of the semiaxes, i.e., of the eigenvalues
of the matrix $\Sigma$. If we write 
\begin{equation}
\label{tranvar}
\Sigma = T^t Q T, 
\end{equation} 
we may think of the six independent entries of $\Sigma$ as consisting of the three
eigenvalues together with the three angles needed to specify the
rotation matrix $T$. We may equally regard the potential energy as a
function of $\Sigma$:
\beq
 \label{vofSigma} 
{\cal V}(\Sigma)={\cal W}(Q).
\eeq

We use the total energy for the Hamiltonian function:
\beq
 \label{trueham} 
H = (1/2)\mbox{Tr}\left({\cal M}^t \Sigma ^{-1} {\cal M}\right) 
+ {\cal V}(\Sigma). 
\eeq
That this function depends only on the moments ${\cal M}$ and $\Sigma$
shows that a reduction has been achieved\footnote{This should be
compared with Rosensteel's equation (4), where an extra term,
proportional to the fluid pressure, appears. It is this extra term
that leads to the correct dynamical equations under Rosensteel's bracket.} . Derivatives of the Hamiltonian are given by the formulas
\beq
\frac{\partial H}{\partial \Sigma} = - \frac{1}{2} \Sigma^{-1} \MM
\MM^t \Sigma ^{-1} + \frac{\partial {\cal V}}{\partial \Sigma}
\label{h_sigma}
\eeq
and
\beq
\frac{\partial H}{\partial \MM} = \Sigma^{-1} \MM\,,
\label{h_m}
\eeq
where indices have been suppressed: to get the $ij$ derivative on the
left, one takes the $ij$ entry of the matrix on the right.

We now find, using Rosensteel's bracket relations
(\ref{S_S_lp})-(\ref{M_S_lp}), the equations of motion
\beqan \label{rhameq1}
\dot{\cal M}_{ij} &=& \left( {\cal M}^t \Sigma ^{-1} {\cal M} \right)_{ij} + \frac{\partial {\cal
V}}{\partial \Sigma _{kl}} \left\{ {\cal M}, \Sigma _{kl} \right\}_R
\quad \mbox{and} \\ \dot{\Sigma}&=& {\cal M} + {\cal
M}^t.\label{rhameq2} \eeqan
These represent a dynamical system of dimension eighteen that has a
fifteen-dimensional invariant manifold expressed by the symmetry of
$\Sigma$, and we henceforth restrict considertion to this system of
dimension fifteen.
With the aid of the relation (\ref{M_S_lp}) we may rewrite the first
of these equations as
\beq \label{rhameq3}
\dot{\cal M} = {\cal M}^t \Sigma ^{-1} {\cal M} -2\Sigma
\frac{\partial {\cal V}}{\partial \Sigma}. \eeq

{F}rom the relations $\Sigma _{ij}=T_{ri}Q_{rs}T_{sj}$ (see equation (\ref{tranvar}) above) and the chain
rule, we find that
\[ \frac{\partial {\cal W} }{\partial Q_{ij}} = \frac{\partial {\cal
V} }{\partial \Sigma _{kl}}\frac{\partial \Sigma _{kl}}{\partial
Q_{ij} } =\frac{\partial {\cal
V} }{\partial \Sigma _{kl}}T_{ik}T_{jl}=\left( T \frac{\partial {\cal
V} }{\partial \Sigma} T^t \right)_{ij},\]
where, in the next-to-last term, we have exploited the fact that $T$ and $Q$
may be regarded as independent. Equation (\ref{rhameq3}) therefore
takes the form
\[\dot{\cal M} = {\cal M}^t \Sigma ^{-1} {\cal M} -2\Sigma
T^t \frac{\partial {\cal W} }{\partial Q} T. \]
Next writing ${\cal M}=T^t {\cal N} T$ and $\Sigma = T^t Q T$ we
obtain the equations in the rotating frame:
\beq
 \label{rhameqsrot}
\dot{\cal N} + [{\cal N},\Omega]= {\cal N}^t Q^{-1} {\cal N} -2Q
\frac{\partial {\cal W}}{\partial Q} \quad \mbox{and} \quad \dot{\cal
Q} + [Q,\Omega] = {\cal N} + {\cal N}^t.
\eeq
Here $\Omega = \dot{T}T^t$ is the antisymmetric angular-velocity
matrix and the square bracket is the commutator:
$[A,B]=AB-BA$. At first glance equations (\ref{rhameqsrot}) do not
look like a well-determined dynamical system. The variables ${\cal N},Q$ appear on the left-hand
side but on the right are ${\cal N},Q$ and $\Omega$, so this system is
well-determined only if $\Omega$ is a function of ${\cal N}$ and $Q$. However, the
second equation consists of three differential equations for
$Q_{11},Q_{22},Q_{33}$ and three equations expressing $\Omega = \Omega
({\cal N},Q)$. They therefore indeed represent a dynamical system of
dimension twelve for these variables. The remaining three variables
of the original fifteen define the rotation matrix $T$ and may be recovered if desired from
the equation $\dot{T}=\Omega T$ after the time dependence of $\Omega$
has been found.  Thus we can think of the transformation from $(\Sigma,\mathcal{M})$ to 
 $(Q,T,\mathcal{N})$ as a change of coordinates. 

The equations of motion for the Riemann ellipsoids in their standard form
as given in equation (\ref{eq57}) below likewise represents a
twelve-dimensional system.
We next bring the moment system (\ref{rhameqsrot}) into this standard form.
Recall that the matrix ${\cal N}$ represents the
set of first moments of the momentum $\rho {\bf U}$ where $U=Kx$ is
linear in $x$:
\[ {\cal N}_{ij}= \int \rho x_i U_j\,d^3x = \int \rho x_i
K_{jl}x_l\,d^3x = Q_{il}K_{jl}.\]
Accordingly, we replace the matrix ${\cal N}$ with $K$ through the transformation 
\beq
 \label{NKconversion} 
{\cal N} =QK^t. 
\eeq
The moment equations (\ref{rhameqsrot}) take the forms
\beqan 
\label{Keq}
\dot{K}&+& K^2 -\Omega K + K \Omega = -2\frac{ \partial {\cal W}
}{\partial Q} \quad \mbox{and}\\
\dot{Q} &=& QK^t + KQ^t + \Omega Q - Q \Omega. \label{Qeq2}
\eeqan

We focus our attention first on the second of these equations,
equation (\ref{Qeq2}). Equation (\ref{Qeq2}),
which is unchanged under transposition, may be regarded as six equations
for the nine entries of $K$. We introduce the matrix 
\beq 
\label{Adef} 
A=\mbox{diag}(a_1,a_2,a_3),
\eeq
the matrix of semiaxes, so that $Q=(m/5)A^2$. The diagonal entries of $K$ are easily found 
to be (for example) $K_{11}= \dot{a_1}/a_1$, by virtue of equation (\ref{Q}). 
Among the off-diagonal entries there must be three that are as yet undetermined. 
If we define a matrix $\Lambda $ through the formula
\beq 
\label{K} 
K= \dot{A}A^{-1} + A\Lambda A^{-1} - \Omega,
\eeq
we find that equation (\ref{Qeq2}) is satisfied if and only if the
matrix $\Lambda$ is antisymmetric. This prescribes the nine entries of
$K$ through the three entries of $\dot{A}A^{-1}$, the three
independent entries of $\Omega$,  and the three independent entries of $\Lambda$.
This should be compared with
\cite{chandra}, chapter 4, equation (42), where the same result is
arrived at in a different way.
 
With the choice (\ref{K}) for $K$, we can now express the left-hand
side of equation (\ref{Keq}) in terms of the variables
$A,\dot{A},\Omega, \Lambda$. We find
\beq \label{lhs}
\dot{K} + K^2 + [K,\Omega]=
\left[ \frac{d^2A}{dt^2} + \frac{d}{dt}\left(A \Lambda - \Omega A
\right) + \dot{A}\lambda - \Omega \dot{A} + A \Lambda ^2 + \Omega ^2 A
- 2\Omega A \Lambda\right] A^{-1} ,
\eeq
i.e., the left-hand side of equation (\ref{Keq}) agrees exactly with
that of equation (\ref{eq57}) of Appendix \ref{summary}. The right-hand side of equation
(\ref{Keq}) is diagonal with, for example, the $11$ entry
\beq \label{11} -2\frac{\partial {\cal W}}{\partial Q_{11}}= -(10/m)
\frac{\partial {\cal W}}{\partial a_1^2}=+3mG \frac{\partial {\cal
I}}{\partial a_1^2}= (3/2)mG{\cal A}_1,\eeq where we have used the
definition (\ref{coeffs}) of ${\cal I}$. 

This gives agreement with
equation (\ref{eq57}) with the important exception that the pressure term
is missing.


\section{Dynamical Equations Under the Dirac Bracket}
\label{diracsec}

We address here two aspects of results of the preceding section that
are not wholly satisfactory. One is the apparent need for a
Hamiltonian that is not the total energy as usually defined, and the
other is that the system obtained is not self-contained but needs to
be augmented by the further constraints alluded to above regarding the
density and the divergence. The latter may seem an innocent
requirement since such augmentation is needed also in the
fluid-dynamical derivation as presented in (\cite{riemann}) or
(\cite{chandra}); see also the discussion in Appendix \ref{summary} below. However, the Hamiltonian version, as embodied in the
bracket (\ref{lie-poisson2}) above, incorporates not only the law of
conservation of momentum but also that of conservation of
mass.\footnote{It would also include conservation of energy (or entropy) if we used
the full bracket as given in (\cite{morrison98}).} We should therefore
expect the dynamics to be fully described by a Hamiltonian description
without the need for any augmentation.

Consider the Dirac bracket $\{\cdot , \cdot \}_{RD}$
presented in equations (\ref{S_S_D}), (\ref{M_M_D}),  and
(\ref{M_S_D}), and again employ as the Hamiltonian the total energy
(equation \ref{trueham}). The additional terms added by the Dirac
procedure to the right-hand sides of the bracket relations
provide corresponding additional terms on the right-hand sides of the
dynamical equations. The dynamical equations corresponding to
equations (\ref{rhameq1}) and (\ref{rhameq2}) therefore become (after
a series of tedious but straightforward calculations)
\beqan 
\label{dirac1}
\dot{\cal M}&=& {\cal M}^t \Sigma ^{-1} {\cal M} +
\frac{1}{\mbox{Tr}(\Sigma ^{-1})}  \mbox{Tr}\left( K^2 + 2
\frac{\partial {\cal W}}{\partial Q} \right)I -
\frac{C^2}{\mbox{Tr}\left( \Sigma ^{-1} \right)} \left( \Sigma ^{-1}
{\cal M} + {\cal M}^t \Sigma ^{-1} \right) \quad \mbox{and} \\
\dot{\Sigma}&=& {\cal M} + {\cal M}^t - 
2 \frac{ C^2 }{ \mbox{Tr}\left( \Sigma ^{-1} \right) } I,
\label{dirac2}
\eeqan
where $I$ denotes the unit matrix and $C^2$ is one of the two Casimir
invariants of the Dirac bracket defined in equation (\ref{C1C2defs})
and is therefore a constant of the motion for the preceding dynamical system.
Since it is proportional to the divergence of the velocity field, it
is supposed to vanish, and we choose the initial data so that this is
so; this simplifies the preceding equations.  
Proceeding as in \S \ref{rosensec}, we obtain from these, with the same definition of
$K$ as in equation (\ref{K}) above, the equation
\beq \label{dirac57}
\dot{K} + K^2 + [K,N] = -2\frac{\partial {\cal W}}{\partial Q} +
\left[ \frac{1}{\mbox{Tr}\left(Q^{-1}\right)}\left(K^2 + 2\partial {\cal W}/\partial Q
\right) \right]Q^{-1}.\eeq

This not only has the structure of equation (\ref{eq57}) but also
explicitly provides the expression for the pressure that is otherwise obtained by the
standard fluid-dynamical procedure needed to maintain
the vanishing of the divergence of the velocity field. To see this,
observe that the term $2p_c/\rho$ of equation (\ref{eq57}) is expressed in terms of the
dynamical variables by taking the trace of each side of equation (\ref{eq57}): 
\[ \frac{2p_c}{\rho}\, \mbox{Tr}(A^{-2}) = \mbox{Tr}\left(\dot{K} +
K^2 + [K,\Omega] + 2 \partial {\cal W}/\partial Q \right) =
\mbox{Tr}\left( K^2 + 2 \partial {\cal W}/\partial Q \right).\]
Here we have used the identity (\ref{lhs}), we have used the formula
(\ref{11}), we have observed that
$Tr([K,\Omega]) = 0$, and we have set 
\[ \mbox{Tr}(\dot{K})= \frac{d}{dt}\left( \sum \dot{a}_i/a_i\right)
=\frac{\partial}{\partial t} \mbox{div}U = 0,\]
in accordance with the fluid-dynamical procedure for defining the pressure.
This gives for the pressure term on the right-hand side of equation
(\ref{eq57}) the expression
\beq
\label{Pcon}
\frac{2p_c}{\rho} A^{-2} = \frac{1}{\mbox{Tr}(A^{-2})}\,
\mbox{Tr}\left( K^2 + 2 \partial {\cal W}/\partial Q \right) A^{-2} = \frac{1}{\mbox{Tr}(Q^{-1})}  \,
\mbox{Tr}\left( K^2 + 2 \partial {\cal W}/\partial Q \right) Q^{-1}.
\eeq
The latter is exactly the extra term provided by the Dirac-bracket
formulation and completes the verification that the dynamics given by
the Hamiltonian (\ref{trueham}) under the Dirac bracket is exactly that of
the Riemann ellipsoids.


\section{Discussion}
\label{disc}

Beginning with the Hamiltonian structure of the ideal fluid, we have
shown that the incompressible Riemann ellipsoids are governed by
Hamiltonian equations in which the Hamiltonian function is the total
energy and the constraints of incompressiblity are incorporated into a
nonlinear bracket via the Dirac formalism.  No extraneous constraints
are required in our formulation. Our results are obtained by
introducing a Dirac bracket for the finite-dimensional system of
moment equations governing the motions of the Riemann
ellipsoid, and are related in spirit to work of Nguyen and Turski
(\cite{turski99}), who formally introduce a
Dirac bracket for the purpose of achieving a Hamiltonian formulation
of the full, infinite-dimensional system of incompressible Euler
equations.

Below we make some additional remarks about constraints.  In
particular, we show that a formulation of Lewis
et al.\ (\cite{lewis87}) for a free boundary liquid, which enforces the incompressibility constraint by requiring
divergence free functional derivatives, gives the correct equations
for a self-gravitating liquid mass.


\subsection{A Bracket for a Free-Boundary Problem}
\label{other}

Lewis et al.\ (\cite{lewis87}) have proposed the following  bracket for a liquid with uniform density and a free boundary: 
\begin{equation}
\label{lmmr_bracket}
\left\{F,G\right\}= \int _D \frac{\delta F}{\delta \mathbf{v}}\cdot
\left( \frac{\delta G}{\delta \mathbf{v}} \times \mathbf{\omega}
\right) \, d^3x + \int _{\partial D} \left(\frac{\delta F}{\delta
\sigma} \frac{\delta G}{\delta \phi} - \frac{\delta F}{\delta
\phi} \frac{\delta G}{\delta \sigma} \right)\, d^2x,
\end{equation}
where $\mathbf{\omega}= \mbox{curl}\, \mathbf{v}$ and the variations have the following meanings. The functionals $F$
and $G$ depend on the velocity field $\mathbf{v}$ in the domain $D$
and also on a variable $\sigma$ determining the instantaneous shape of
the boundary and defined as follows. The distance $\Delta \sigma$ is
the amount that some point $\mathbf{x}$ on $\partial D$ moves
normal to itself in the time interval $\Delta t$. Therefore 
$\sigma _t = \hat{\mathbf{n}} \cdot \mathbf{v}$. This is the local evolution equation for the motion of the surface normal to itself. The variable $\sigma$ is therefore a
function of surface coordinates on $\partial D$ and of time. The variational derivative $\delta
F/\delta \mathbf{v}$ is clear, but what is less obvious is the
requirement that it, like the velocity $\mathbf{v}$, be solenoidal:
\begin{equation}
\label{solvar}
\mbox{div}\, \frac{\delta F}{\delta \mathbf{v}} =0.
\end{equation}
The remaining functional derivative is given by the formula
$\delta F/\delta \phi = \hat{\mathbf{n}} \cdot \delta F/\delta \mathbf{v}$. 
It is evaluated only on the boundary and is not an independent variation
but depends on  $\delta F/\delta \mathbf{v}$.


In their paper, Lewis et al.\  show how  this bracket yields the 
equations of motion for a liquid drop held
together by surface tension.  We
now verify that it does the same if surface tension in the Hamiltonian is replaced by
self-gravitation. The Hamiltonian is then 
$H[\mathbf{v},\sigma]=T[\mathbf{v},\sigma] + W[\sigma]$, 
where
\[ T[\mathbf{v},\sigma]=\int _D (1/2) |\mathbf{v}|^2 \, d^3x \quad 
\mbox{and} \quad  W[\sigma]=-(1/2)\int _D V(x)\, d^3x,\  V(x)=\int
_D\frac{d^3y}{|\mathbf{x} - \mathbf{y}|}.
\]
The dependence on $\sigma$ arises because  the domain $D$ depends on the
shape of the boundary. Straightforward calculations show that
\begin{equation}
\label{hamvar}
\frac{\delta H}{\delta \mathbf{v}}=\mathbf{v},\quad  \frac{\delta
H}{\delta \phi} = \hat{\mathbf{n}}\cdot \mathbf{v}, \; \frac{\delta
H}{\delta \sigma } = (1/2)|\mathbf{v}|^2 - V(x).
\end{equation}
These variations have been made without explicitly imposing the solenoidal
constraint (\ref{solvar}), but note that $\delta H/\delta \mathbf{v}$
satisfies this constraint anyway by virtue of the solenoidal character
of $\mathbf{v}$.
Therefore
\begin{eqnarray*}&& \{F,H\}=\int _D  \frac{\delta F}{\delta \mathbf{v}}
\cdot \left( \mathbf{v} \times \mathbf{\omega}\right) \, d^3x + \int
_{\partial D}  \hat{\mathbf{n}}\cdot
\mathbf{v} \frac{\delta F}{\delta \sigma} \, d^2x - \int
_{\partial D} \hat{\mathbf{n}}\cdot  \frac{\delta F}{\delta \mathbf{v}}
 \left( (1/2)|\mathbf{v}|^2 + V(x)\right) \, d^2x \\
&& = \int _D \frac{\delta F}{\delta \mathbf{v}} \cdot \left(
-\mathbf{v} \cdot \nabla \mathbf{v} + \nabla V(x) \right) \, d^3x 
+ \int _{\partial D}\frac{\delta F}{\delta \sigma} \hat{\mathbf{n}}\cdot \mathbf{v} \, d^2x\,, 
\end{eqnarray*}
where we have used the fact that the divergence of $\delta
F/ \delta \mathbf{v}$ vanishes and a standard vector identity.

On the other hand,
\[ F_t = \int _D  \frac{\delta F}{\delta \mathbf{v}} \cdot \mathbf{v}_t
\, d^3x + \int _{\partial D} \frac{\delta F}{\delta \sigma} \sigma _t
 \, d^2x .\]
Hamilton's equations hold if and only if $F_t=\{F,H\}$ for all
functionals $F$. Comparing the expressions for the two quantities we
see that we must have $\sigma _t = \hat{\mathbf{n}}\cdot \mathbf{v}$,
expressing the free-boundary condition. The equality of the two
integrals multiplied by $\delta F/\delta \mathbf{v}$ does not
guarantee the equality of their coefficients because $\delta F/\delta
\mathbf{v}$ is not entirely arbitrary but in the Lewis et al.\ formulation 
must be constrained by the solenoidal condition: if $p(x)$ is any function on $D$ vanishing on $\partial D$, $\int _D  \delta F/\delta \mathbf{v} \cdot \nabla p \,
d^3x=0$. 
Thus the equality of $F_t$ with $\{F,H\}$ implies the correct equation of motion, 
$\mathbf{v}_t=- \mathbf{v} \cdot \nabla \mathbf{v} - \nabla p - \nabla V(x)$, 
where $p$ is a scalar vanishing on $\partial D$. 


In principle one should next check whether the moments $(\Sigma,{\cal M})$ effect a
reduction with the Lewis et al.\  procedure. Because  we know that the Hamiltonian depends only on these moments, this amounts to checking that they are closed under the brackets. With the definitions of (\ref{defmom1}) and (\ref{defmom2})
we find for the variational derivatives, on ignoring the solenoidal constraint,
\[ \frac{\delta {\cal M}_{ij}}{\delta v_k}=x_i \delta _{jk}\,, \quad
 \frac{\delta {\cal M}_{ij}}{\delta \sigma} = x_iv_j\,, \quad 
 \frac{\delta \Sigma_{ij}}{\delta v_k}=0\,,  \quad
\mbox{and} \quad  \frac{\delta \Sigma_{ij}}{\delta \sigma} = x_ix_j.
\]
It is seen that $\delta \Sigma/\delta \mathbf{v}$ satisfies this
constraint, but $\delta {\cal M}/\delta \mathbf{v}$ does not. This can
be rectified by restricting also the variations $\delta \mathbf{v}$ to
be solenoidal, thereby modifying the expression for $\delta {\cal
M}/\delta \mathbf{v}$ by the addition of a certain gradient. Carrying
this out, checking algebraic closure, and verifying the equations of
motion of the Riemann ellipsoid would require calculations of a
length and difficulty similar to those already carried out in this paper and
we do not record these here.

Fasso and Lewis (\cite{fasso01}) have given an alternative Hamiltonian
formulation, not for fluid dynamics, but explicitly for the equations
governing the Riemann ellipsoids.



\subsection{The Nature and Number of Incompressibilty Constraints}
\label{nature}

The Dirac procedure requires an even number of constraint functions
and we have used two. It might be surmised that the goal of
introducing incompressibility would require only one constraint,
$\mbox{div}\,v =0$, and that the imposition of a second is an artifice
needed in order to use the Dirac procedure. This is not so.

It is easiest to see this in the special context of the Riemann
ellipsoids. In equation (\ref{eq57}) there are two extra parameters,
$p_c$ and $\rho$, that need to be defined in order to make the system
determinate. One of these is achieved by simply declaring $\rho$ to be
a fixed constant. The second is achieved by taking the trace of either
side of the equation and setting 
$$\frac{\partial}{ \partial t} \mbox{div}\,v = \frac{d}{dt} \, \left(\sum \dot{a}_i/a_i \right)
=0 ,$$
thereby defining $p_c$ as a function of the velocity field. 
This definition of $p_c$ ensures that the preceding equation will hold
for all $t$ and therefore that $\sum \dot{a}_i/a_i =0$ for all $t$ if
this is chosen to be true at the initial instant. Our choice of
two invariants for the Dirac bracket corresponds precisely to these
choices.

In a more general fluid-dynamical framework in which velocity and
density vary with position, the imposition of the
constraint $\mbox{div} \, v =0$ is not a single constraint, but an
infinite family of constraints indexed by the position vector
$x$. Once imposed, it implies by virtue of mass-conservation equation (\ref{euler2})
that $D\rho/Dt = 0$, where $D/Dt = \partial /\partial t + v\cdot
\nabla$ represents the convective derivative. This means the initial
values of the density are convected by the velocity field and
necessitates the imposition of a second family of conditions, namely
those determining the density at the initial instant of time.

\subsection{Invariants}
\label{invariants}

Notice that the mass $m$ is the zeroth moment of the density distribution
and an algebra reduction can be constructed for it.  It is a
Casimir invariant and, as one would expect, so is the first moment 
(the center-of-mass position).  
By restricting attention to the quadratic moments of the density we sit
on the symplectic leaf of constant mass and center-of-mass position.
In the algebra we have constructed, aside from the Casimirs that we
have introduced, there is one more.

Rosensteel \cite{rosensteel88} shows that the magnitude of the Kelvin circulation vector
\beq
\Gamma^2 \equiv 
{\rm Tr}\left[\Sigma^{-1}\MM\Sigma\MM^t - \MM\MM\right]
\label{kelvin}
\eeq
is a Casimir for the algebra
$(gcm\lp 3\rp)$ and it remains so for the present
algebra\footnote{This refers to the system in the rotating frame. When
the equations of motion are written in the inertial frame, it is
possible to identify a three-component vector of circulation, each of
whose components is separately conserved.}.  
That it is a Casimir for Rosensteel's unconstrained algebra shows that
its validity does not depend on incompressiblity.
The angular momentum, $\epsilon_{ijk}\MM_{jk} $ is not a Casimir for this
algebra, but is conserved by the choice of Hamiltonian.


\appendix



\section{Summary of the Equations Governing Riemann Ellipsoids}
\label{summary}

We provide a summary of the basic equation governing the motion of a
self-gravitating, liquid
ellipsoid of spatially uniform density $\rho$ and semiaxes
$a_1,a_2,a_3$ with a velocity field depending linearly on the
cartesian coordinates. A full description is in \cite{chandra}, Chapter 4.

Relative to a rotating reference frame in which the cartesian
coordinates $x$ are aligned with the principal axes of the ellipsoid,
fluid motions are allowed that have the form
\beq 
\label{u_rep}
 u(x)=\left(\dot{A} + A \Lambda \right)A^{-1}x 
\eeq
where $A=\mbox{diag}(a_1,a_2,a_3)$ and $\Lambda$ is an antisymmetric
matrix. $A$ and $\Lambda$ are in general time-dependent, but the full
spatial dependence of $u$ is that of linearity in $x$, as explicitly
expressed in this equation. The rotation rate of this rotating
frame is expressed via a second antisymmetric matrix $\Omega$, and the
dynamical equations governing the time evolution of the variables
$A,\Omega,\Lambda$ may be written as (cf. \cite{chandra}, Chapter 4,
equation 57)
\beq 
\label{eq57}
\left[ \frac{d^2A}{dt^2} + \frac{d}{dt}\left(A \Lambda - \Omega A
\right) + \dot{A}\lambda - \Omega \dot{A} + A \Lambda ^2 + \Omega ^2 A
- 2\Omega A \Lambda \right]A^{-1} 
= -\frac{3}{2}mG{\cal A} + \frac{2p_c}{\rho}A^{-2} \,,  
\eeq
where ${\cal A}={\cal A}(A)=\mbox{diag}({\cal A}_1,{\cal A}_2,{\cal A}_3)$
represents the coefficients in the self-gravitational potential
\beq 
\label{gravpot} 
V(x) = \frac{3}{4}mG\left( {\cal I} - \sum
_{i=1}^3 {\cal A}_i x_i^2 \right)\,,
\eeq
which is valid inside the ellipsoid. These coefficients are determined by the semiaxes via the
formulas\footnote{The definitions given here differ by a factor
$a_1a_2a_3$ from those given in (\cite{chandra})}. 
\beq 
\label{coeffs} 
{\cal I}= \int _0^\infty \frac{du}{\Delta (u)},\; {\cal A}_i = \int _0^\infty
\frac{du}{\left(a_i^2 + u\right)\Delta (u)}, \; \mbox{where} \; \Delta (u) = \sqrt{(a_1^2 + u)(a_2^2 +u)(a_3^2+u)}.
\eeq
The scalar $p_c$ is the pressure at the center $x=0$.

The system (\ref{eq57}) consists of twelve first-order equations in the twelve
unknowns of $A,\dot{A},\Omega \mbox{ and }\Lambda$ in which $\rho$ and
$p_c$ appear as parameters. It arises from 
equation (\ref{euler1}) only, i.e., from the imposition of the law of
conservation of momentum only. It must be augmented by further
information in order to render it determinate. For incompressible
flow, two conditions are imposed that are consistent with  equation
(\ref{euler2}) of mass conservation: the density\footnote{Or
alternatively the product $a_1a_2a_3$.} is set equal to a constant
(which is therefore excluded from the list of variables) and the
solenoidal condition $\sum \dot{a_i}/a_i =0$ is imposed.
One can then express $p_c$ in terms of the 
dynamical variables $A,\dot{A},\Omega ,\Lambda$ by taking the trace of
each side of equation (\ref{eq57}) and putting $\frac{d}{dt}\sum
\dot{a}_i/a_i = 0$; then one has twelve equations in twelve unknowns in
which the solenoidal condition $\sum \dot{a}_i/a_i = 0$ is preserved
by virtue of the choice of $p_c$ together with the initial
data\footnote{Alternatively one can eliminate $p_c$ from the system
and achieve a system of ten equations in ten unknowns.}.

\section{The Hybrid Coordinate systems}
\label{hybrid}

The transformation  of equations (\ref{dirac1}) and (\ref{dirac2})
to the equations governing the dynamics of $(Q,T,\mathcal{N})$ was 
demonstrated in \S \ref{rosensec}, and  their equivalence to Riemann's equations  of  
(\ref{eq57}) with(\ref{Pcon}) was demonstrated in \S \ref{diracsec}.  Thus Riemann's equations are simply the moment equations as we have derived them with velocities
and coordinates resolved in a reference frame rotating with the 
body of the ellipsoid.  

In fact, there are four reference frames of interest.  The {\it first}, with variables $(\Sigma,\mathcal{M})$,  uses velocities measured in the inertial frame
and resolved  along axes in the inertial frame while the {\it fourth}, with variables $(Q,\mathcal{N})$,  supressing the dependence on the rotation matrix, measures and resolves velocities in the rotating frame (where Riemann's 
equations live).  There is also a {\it second},  hybrid frame, with variables $(\Sigma,\tilde {\mathcal{M}})$, where velocities are measured in the inertial frame but resolved along axes in the rotating frame and a {\it third},  hybrid frame, with variables $(Q,\tilde{\mathcal{N}})$,where velocities are measured in the rotating frame but resolved along axes in the inertial frame.  We  present the transformation to these frames here.

For the fourth frame, we showed in \S \ref{rosensec} that with  $\Sigma = T^t QT$ and $\NN = T \MM T^t$ the equations of motion for $(\Sigma,\MM)$, (\ref{dirac1}) and (\ref{dirac2}),  become
\beq
\dot{\cal Q} = [\Omega,Q] + {\cal N} + {\cal N}^t
\quad \mbox{and} \quad 
\dot{\cal N} = [\Omega,{\cal N}] + {\cal N}^t Q^{-1} {\cal N} + {\cal F}.
\label{NQagain}
\eeq
where ${\cal F}$ represent pressure and forcing terms which transform
in a straightforward fashion.  

In the third frame the velocities are resolved along the inertial
frame coordinates but are measured along some rotating frame.  
At the outset  there is no need to bias this frame by requiring it
to be the frame rotating with the body so we can consider an 
arbitrary angular velocity vector $\omega$ such that
${\bf u}^{Rot} = {\bf u}^{Inert} - {\bf \omega} \times {\bf x}$. 
So,  defining $\tilde{\MM}_{ij} = \int \rho x_i u^{Rot}_j \; d^3 x$
we find
\beq 
\tilde{\MM}_{ij} =  \MM_{ij} - 
\int \rho x_i \epsilon_{jkl} \omega_k x_l \; d^3 x
= \MM_{ij} - \Sigma_{il} \tilde{\Omega}_{lj} \,, 
\eeq
where $\tilde{\Omega}_{lj} = \epsilon_{ljk} \omega_k$.  
Therefore the dynamics of $\Sigma$ and $\tilde{\MM}$ are
governed by
\beqan
\dot{\Sigma} &=& \tilde{\MM} + \tilde{\MM}^t + \Sigma \tilde{\Omega}
-\tilde{\Omega} \Sigma
\label{sig_dot_rot} \\
\dot{\tilde{\MM}} &=& \tilde{\MM}^t \Sigma^{-1} \tilde{\MM} -
\tilde{\Omega} \tilde{\MM} - \tilde{\MM}\tilde{\Omega} - 
\Sigma \tilde{\Omega}\tilde{\Omega} - \Sigma\dot{\tilde{\Omega}} + 
{\cal F}.
\label{m_tilde_dot}
\eeqan
So far, $\tilde{\Omega}$ can be a completely arbitrary, prespecified
function of time.  The terms on the right had side of (\ref{m_tilde_dot})
represent advection, Coriolis, centripetal, Euler and
external forces, respectively.  If we choose a frame to coincide with
the body of the ellipsoid, then $\Sigma$ must be diagonal 
and, in this manner, $\tilde{\Omega}$ is determined.

The equations for moments completely specified in the rotating reference
frame can be arrived at by either conjugating 
(\ref{sig_dot_rot}) and  (\ref{m_tilde_dot}) with an orthogonal matrix
or by shifting the velocity in (\ref{NQagain}).  We
shall perform both. Defining  $\tilde{\NN} = \NN -  Q \Omega$ and inserting into (\ref{NQagain}) gives easily 
\beqan
\dot{Q}&=& \tilde{\NN}^t + \tilde{\NN}
\label{d_dot2}
\\
\dot{\tilde{\NN}}&=& \tilde{\NN}^t Q^{-1}\tilde{\NN} - 2 \tilde{\NN} \Omega - 
Q\Omega\Omega - Q \dot{\Omega} + {\cal F}.
\label{n_tilde_dot}
\eeqan
Alternatively, using   $Q = T \Sigma T^t$ and $\tilde{\NN} = T \tilde{\MM} T^t$, substituting  in
(\ref{sig_dot_rot}) and  (\ref{m_tilde_dot}), and identifying
$\Omega =  T \tilde{\Omega} T^t$    gives again  (\ref{d_dot2})  and (\ref{n_tilde_dot}).  Note,  with the above definition of $\tilde{\Omega}$,  defining $\tilde{T}$
by $\dot{\tilde{T}} = -  \tilde{\Omega} \tilde{T}$,   results in $\tilde{T}^t=T$.

\section{The Jacobi Identity for General Dirac Brackets}
\label{jacobi}

It is known (cf.\  \cite{sudarshan}) that a Dirac bracket based on a
canonical bracket satisfies the Jacobi identity and therefore provides
a valid bracket. To our knowldge there is no explicit corresponding proof in
the literature for the case when the original bracket is more general,
i.e., not necessarily canonical. We provide that proof here.  

We must show that
\beq
\{\{F,G\}_D,H\}_D + {\rm \; cyclic \; permutations} = 0
\label{dirac_jacobi}
\eeq
for all $F,\,G,\,H$ and any invertible $\omega$. Therefore
\beqan
\{\{F,G\}_D,H\}_D &=& \{\{F,G\},H\}_D - 
                \{  \{F,C_{\mu}\} \,\omega^{-1}_{\mu \nu} \,
                \{C_{\nu},G\} , H \}_D 
\nonumber \\
&=& \{\{F,G\},H\} - 
\{ \{F,G\},C^{\mu} \} \,\omega^{-1}_{\mu \nu} \, \{C^{\nu}, H\}
\nonumber \\
&& -\{ \{F, C^{\mu} \} \,\omega^{-1}_{\mu \nu} \, \{C^{\nu}, G \}, H \}
\nonumber \\
&& + \{ \{ F, C^{\alpha} \} \,\omega^{-1}_{\alpha \beta} \, 
\{C^{\beta}, G \}, C^{\mu} \} \,\omega^{-1}_{\mu \nu} \, 
\{C^{\nu}, H \} 
\nonumber 
\eeqan
where the subscipts on the right hand side have all been dropped
in the second line since it is unambiuously written in terms of the 
Lie-Poisson bracket.  Upon cyclic permutations, the first
term will cancel due to the Jacobi identity which holds for the
Lie-Poisson bracket, so we can dispose of it immediately.  
Using the Leibnitz rule, the left hand side of 
(\ref{dirac_jacobi}) becomes 
\beqa
&=& -\{ \{F,G\},C^{\mu} \} \,\omega^{-1}_{\mu \nu} \, \{C^{\nu}, H\}
 -\{ \{F, C^{\mu} \}, H \} \,\omega^{-1}_{\mu \nu} \, \{C^{\nu}, G \}
\\
&&
 - \{F, C^{\mu} \} \,\omega^{-1}_{\mu \nu} \,\{ \{C^{\nu}, G \}, H \}  
- \{F, C^{\mu} \} \{ \,\omega^{-1}_{\mu \nu} \, , H \} \{C^{\nu}, G \}
\\
&& + \{ \{ F, C^{\alpha} \}, C^{\mu} \}  \,\omega^{-1}_{\alpha \beta} \, 
\{C^{\beta}, G \}\,\omega^{-1}_{\mu \nu} \, 
\{C^{\nu}, H \} 
\\
&& + \{ F, C^{\alpha} \} \,\omega^{-1}_{\alpha \beta} \, 
\{ \{C^{\beta}, G \}, C^{\mu} \} \,\omega^{-1}_{\mu \nu} \, 
\{C^{\nu}, H \} 
\\
&& + \{ F, C^{\alpha} \} \{ \,\omega^{-1}_{\alpha \beta} \, 
, C^{\mu} \} \{C^{\beta}, G \} \,\omega^{-1}_{\mu \nu} \, 
\{C^{\nu}, H \} + {\rm \; c.p.'s}.
\eeqa
The $\omega^{-1}$ term can be pulled out of the bracket in all
of the terms by recognizing the relation
\beqan
\{ \omega^{-1}_{\mu \nu}, F\} &=& -\omega^{-1}_{\mu \alpha}
\omega^{-1}_{\beta \nu} \{ \omega_{\alpha \beta}, F \}  \\
&=&  -\omega^{-1}_{\mu \alpha}
\omega^{-1}_{\beta \nu} \{ \{C^{\alpha},C^{\beta} \}, F \} .
\eeqan
The first three terms and their permutations cancel due to
the Jacobi Identity as do the second three terms and 
their permutations.  Finally, the last term and its permutations cancel 
amongst themselves due to the Jacobi identity.  In this way, it
can be shown that the Dirac bracket defines a  Lie algebra with an even number  of Casimirs more than the  original algebra for {\em any} bracket.

 
\section{Non-Persistence of Invariants}
\label{example}

The Dirac bracket construction ensures that the existence of the
Lie-Dirac invariants.  However, if there exist other dynamical
invariants of the unconstrained system, i.e.\ invariants that commute
with the Hamiltonian under the unconstrained bracket, canonical or
Lie-Poisson, then there is no reason that these invariants will remain
invariants under the Dirac bracket dynamics.  Here we give
an example where dynamical invariance is lost.

Consider an  $N$-body type of system with a Hamiltonian of the form
\beq
H(p,q)=\sum_{i=1}^{N} \frac{{p}_i^2}{2} + \mathcal{V} 
=\sum_{i=1}^{N} \frac{{p}_i^2}{2} + \sum_{i,j=1}^{N} V(x_i-x_j)\,, 
\eeq
where $V(x_i-x_j)=V(x_j-x_i)$, and dynamics generated under the  canonical Poisson bracket, 
\beq
\{f,g\}= \sum_{i=1}^{N}
\left(\frac{\partial f}{\partial x_i}\frac{\partial g}{\partial p_i} 
- \frac{\partial g}{\partial x_i}\frac{\partial f}{\partial p_i}
\right)
\,. 
\eeq
This system   conserves the total momentum $P=\sum_{k=1}^{N} p_i$, 
as is easily shown.

Now, suppose we constrain away one of the degrees of freedom, by choosing 
\beq
C^1=x_1\quad\quad {\rm and }\quad\quad C^2=p_1\,
\eeq
which results in the following Dirac bracket:
\beq
\{f,g\}_D=  \sum_{i=2}^{N}
\left(
\frac{\partial f}{\partial x_i}\frac{\partial g}{\partial p_i} 
- \frac{\partial g}{\partial x_i}\frac{\partial f}{\partial p_i}
\right)
\eeq
Thus under the constrained dynamics
\beq
\dot P= \frac{\partial \mathcal{V} }{\partial x_1} \neq 0\,.
\eeq
We lose Newton's third law because reaction forces are nulled out by the constraint.


\section*{Acknowledgments}
PJM  was supported by the US Department of Energy Contract No.~DE-FG03-96ER-54346 and JAB was supported by an NSF VIGRE Postdoctoral fellowship while at Rensselaer Polytechnic Institute. 


\bibliographystyle{elsarticle-num}

\bibliography{pmbib}

\begin{thebibliography}{10}
\expandafter\ifx\csname url\endcsname\relax
  \def\url#1{\texttt{#1}}\fi
\expandafter\ifx\csname urlprefix\endcsname\relax\def\urlprefix{URL }\fi
\expandafter\ifx\csname href\endcsname\relax
  \def\href#1#2{#2} \def\path#1{#1}\fi

\bibitem{morrison98}
P.~J. Morrison, Hamiltonian description of the ideal fluid, Rev.\ Mod.\ Phys.
  70~(2) (1998) 467--521.

\bibitem{dirichlet}
G.~L. Dirichlet, Untersuchungen \"uber ein problem der hydrodynamik, J. Reine
  Angew. Math. 58 (1860) 181--216.

\bibitem{riemann}
B.~Riemann, Untersuchungen \"uber die bewegung eines fl\"ussigen gleichartigen
  ellipsoides, Abh. d. K\"onigl. Gesell. der Wis. zu G\"ottingen 9 (1860)
  3--36.

\bibitem{chandra}
S.~Chandrasekhar, {E}llipsoidal {F}igures of {E}quilibrium, Dover, New York,
  1987.

\bibitem{rosensteel88}
G.~Rosensteel, Rapidly rotating nuclei as {R}iemann ellipsoids, Ann.\ Phys. 186
  (1988) 230--291.

\bibitem{morrison80}
P.~J. Morrison, J.~M. Greene, {N}oncanonical {H}amiltonian density formulation
  of hydrodynamics and ideal magnetohydrodynamics, Phys.\ Rev.\ Lett. 45 (1980)
  790--793.

\bibitem{turski99}
S.~Nguyen, L.~A. Turski, Canonical description of incompressible fluid: {D}irac
  brackets approach, Physica A 272 (1999) 48--55.

\bibitem{lewis87}
D.~Lewis, J.~Marsden, T.~Ratiu, Stability and bifurcation of a rotating planar
  liquid drop., J.\ Math.\ Phys. 28 (1987) 2508--2515.

\bibitem{fasso01}
F.~Fasso, D.~Lewis, Stability properties of the {R}iemann ellipsoids, J. Rat.
  Mech. Anal. 158 (2001) 259--292.

\bibitem{sudarshan}
E.~C.~G. Sudarshan, N.~Makunda, Classical Dynamics : A Modern Perspective, John
  Wiley {\&} Sons, New York, 1974.

\end{thebibliography}
\end{document}